\documentclass[conference]{IEEEtran}

\usepackage{graphicx}
\usepackage{amsmath}
\usepackage{amssymb}

\newcommand{\s}[1]{\mathcal{#1}}
\newcommand{\br}{\ensuremath{\textbf{r}}}

\begin{document}

\newpage \onecolumn
This manuscript is the accepted version of the following paper published by the IEEE:\\
``Opti (c, m) al: Optical/Optimal Routing in Massively Dense Wireless Networks'' \\
R. Catanuto, S. Toumpis, and G. Morabito \\
Proc. INFOCOM 2007, Anchorage, AL, May 2007.\\
DOI: 10.1109/INFCOM.2007.122   \\
IEEE Xplore link: http://ieeexplore.ieee.org/xpl/articleDetails.jsp?arnumber=4215704 \\

The following copyright notice applies:\\
" \copyright 2007 IEEE. Personal use of this material is permitted. Permission from IEEE must be obtained for all other uses, in any current or future media, including reprinting/republishing this material for advertising or promotional purposes, creating new collective works, for resale or redistribution to servers or lists, or reuse of any copyrighted component of this work in other works."

\title{{\tt Opti\{c,m\}al}: Optical/Optimal Routing in Massively Dense Wireless Networks \thanks{R. Catanuto is with the Dipartimento di Matematica e Informatica, University of Catania. S. Toumpis is with the Department of Electrical and Computer  Engineering, University of Cyprus. G. Morabito is with the Dipartimento di Ingegneria Informatica e delle Telecomunicazioni, University of Catania. Research is supported by the European Network of Excellence CRUISE.}}
\author{\authorblockN{Roberto Catanuto, Stavros Toumpis, Giacomo Morabito\\}}
\maketitle
\thispagestyle{empty}

\begin{abstract}
We study routing for massively dense wireless networks, i.e., wireless networks  that contain so many nodes that, in addition to their usual \emph{microscopic} description, a novel \emph{macroscopic} description becomes possible. The macroscopic description is not detailed, but nevertheless contains enough information to permit a meaningful study and performance optimization of the network. Within this context, we continue and significantly expand previous work on the analogy between optimal routing and the propagation of light according to the laws of Geometrical Optics. Firstly, we pose the analogy in a more general framework than previously, notably showing how the eikonal equation, which is the central equation of Geometrical Optics, also appears in the networking context. Secondly, we develop a methodology for calculating the cost function, which is the function describing the network at the macroscopic level. We apply this methodology for two important types of networks: bandwidth limited and energy limited. 

\end{abstract}
\section{Introduction}

\subsection{Trajectory Based Forwarding and Massively Dense Networks}

In most routing protocols for wireless networks, each route is described in terms of the sequence of nodes that a packet must follow from its source to its destination. The node sequence is either determined at the source (for source driven protocols) or along the way to the destination (for table driven protocols)~\cite{perkins1}. However, in both cases, the network often suffers from the \emph{scalability problem}: As the number of nodes in the network increases, the number of nodes participating in a typical route also increases, and so the route becomes more susceptible to node mobility and channel fluctuations. Therefore, more and more bandwidth must be used to fix errors and keep the routes updated~\cite{santinavez1}. 

In order to solve the scalability problem, the authors of \cite{niculescu1} have endeavored to turn the large number of nodes into an advantage. Notably, they consider a \emph{dense} network, and proceed to propose routing by means of \emph{Trajectory Based Forwarding} (TBF). The network is dense in the sense that it consists of a very large number of nodes so that each node typically has within its communication range several neighboring nodes, that it can use to forward packets toward all directions. In this setting, the route is specified not in terms of a list of nodes, to be visited consecutively on the way to the destination, but rather as a continuous curve, called the \emph{trajectory}, that starts at the physical location of the source and ends at the physical location of the destination. When the packet arrives at an intermediate node, that node forwards the packet to a node that happens, at that time, to be further ahead along the trajectory of the packet, according to a greedy forwarding rule. As continuous curves can be described compactly, and there is no need to specify beforehand the nodes that participate in the route or maintain routing tables, TBF scales with the network size much better than traditional routing protocols. TBF may be thought of as a generalization of geographic routing \cite{giordano1}, with which the trajectory is the straight line connecting the source and the destination.

The concept of dense networks took a significant evolutionary step forward with the introduction of \emph{massively dense} networks \cite{jacquet1, kalantari2,  toumpis17,toumpis20, sirkeci1, hyytia1}. These networks consist of such a large number of nodes, that we can take a novel \emph{macroscopic} view (i.e., bird's eye view) of the network, in addition to the more traditional \emph{microscopic} view, and approximate the various microscopic quantities of interest with concise macroscopic equivalents. 

The precise microscopic and macroscopic quantities considered depend on the particular problem at hand. Typical examples of microscopic quantities are node positions, data transmission rates at individual relaying nodes, and the data creation rates at individual source nodes. Their corresponding macroscopic quantities are the density of nodes (a scalar measured in nodes per square meter), the flow of traffic (a vector measured in bps per meter), and the density of data creation at the sources (measured in bps per square meter), all functions of the location $(x,y)$. 

Focusing on the macroscopic and ignoring the microscopic quantities involves losses in describing the network. For example, if we describe the network deployment in terms of a node density function, we no longer keep track of the particular placements of individual nodes. On the other hand, the macroscopic quantities retain sufficient detail so as to capture the fundamental properties of the network, and permit the derivation of meaningful and insightful results that describe adequately the behavior of the network. 

\subsection{Routing by Analogy to Geometrical Optics}

Of particular interest to our work is the line of investigation initiated in \cite{jacquet1}, and in which the term `massively dense' first appeared. There, the objective is to find the route between a source and a destination that requires the minimum number of hops. The network topology is described in terms of the spatial node density function. It is assumed that communication is restricted between nearest neighbors, so that in areas of higher node density each hop corresponds to a smaller physical distance. As a result, packets should try to steer away from high node density areas. The author shows that, in the massively dense regime, the optimal trajectory satisfies the same set of equations that are satisfied in Geometrical Optics by a ray of light that travels from the source to the destination, assuming the massively dense network is substituted by a properly specified inhomogeneous optical medium. Therefore, we can use the theory, tools, and intuition that exist in Geometrical Optics for solving a problem in networking. 

Despite the high degree of originality of \cite{jacquet1}, the scope of that work is limited: The aim is to find the trajectory from the source to the destination that uses the minimum number of hops, assuming communication is restricted between nearest neighbors. However, in wireless networks communication is typically not limited between nearest neighbors, but rather between nodes that are close enough to hear each other's transmissions. Furthermore, minimizing the number of hops of a route is only one possible objective, very frequently not the most relevant: other objectives are the minimization of delay, bandwidth, energy, etc. 

These limitations motivated the work in \cite{catanuto1}. There, it is assumed that each transmission has a fixed energy cost, and nodes can communicate with each other as long as they are within some maximum distance from each other. The aim is to find the route that uses the least amount of energy. The authors show that this setting actually favors the routing of packets through areas of high node density. Again, they show that in the massively dense regime, the optimal routes satisfy the same set of equations as those satisfied by a ray of light traveling through a properly defined inhomogeneous optical medium. In addition, they place the problem in the context of TBF: each ray of light is a trajectory that the packets must follow according to a trajectory based forwarding rule. (The details of how forwarding is performed on the microscopic level were ignored in \cite{jacquet1}.)

Here, we continue on the line of investigation initiated in \cite{jacquet1} and continued in \cite{catanuto1}, making a number of contributions in two main directions. Firstly, we pose the analogy between Geometrical Optics and massively dense networks in a more general framework, that permits its use under a wider variety of problems. Notably, we specify the networking equivalent of the eikonal equation of Optics. Our framework allows to identify the optimal trajectory in two relevant contexts: a simple routing problem, in which information must be transferred by a given source to a given destination, and a distributed routing problem, in which the source or the destination can be any from a set of known nodes. Observe that the second problem cannot be handled with the frameworks considered in \cite{jacquet1} and \cite{catanuto1}.

Secondly, we develop a methodology for calculating the cost function, which is the key macroscopic quantity for our work, in several scenarios, and we exhibit this methodology in two cases of high relevance: a bandwidth limited network model, and an energy limited network model. 

The rest of this paper is organized as follows. To keep the text self-contained, in Section \ref{Background} we briefly review the basic concepts of Geometrical Optics that we use. In Section \ref{MacroscopicModel} we introduce a macroscopic model of the network which formalizes the analogy between Geometrical Optics and the problem of routing in massively dense wireless networks. The macroscopic model needs the specification of an appropriate cost function to be derived using a microscopic model of the network, and so in Section \ref{MicroscopicModel} we derive the cost function for a bandwidth limited model and an energy limited model. The proposed analogy is assessed in Section \ref{Simulations} where simulation results show that the trajectories derived through the proposed analogy closely follow the route obtained by applying shortest path routing algorithms such as the Bellman-Ford algorithm. Finally, open research issues are given in Section \ref{Conclusions}.

\section{Background on Geometrical Optics} 
\label{Background}

In this section we provide a brief overview of Geometrical Optics~\cite{born1, hecht1}, and in particular those aspects that are related to our work. As our networks are 2-dimensional, we discuss Optics in a 2-dimensional setting.

Geometrical Optics describes the propagation of electromagnetic fields when the wavelength is much smaller than the dimensions of the features of the environment or, more technically, in the limit where its wavelength goes to zero. In this approximating case, the laws of propagation can be formulated in the language of geometry, hence the name of this theory. Note that the wavelengths of visible light are on the order of micrometers, and therefore visible light is very well described by Geometrical Optics. 

The emanation of light from a light source is described in terms of the \emph{eikonal function} $S(\br)$ which is measured in meters and equals the time it takes for the light to go from the source to point $\br$, multiplied with the speed of light in free space $c$. Therefore, the eikonal function $S(\br)$ represents the distance that light will cover \emph{in free space}, at the time it takes light to travel, \emph{through the medium}, from the source to the point $\br$. Surfaces of the form $ S(\br)=k>0$ are called \emph{wavefronts}, and represent loci of points at which light, leaving the source at time $0$, will arrive at time $\frac{k}{c}$.

\begin{figure}
\begin{center}
\includegraphics[scale=0.5]{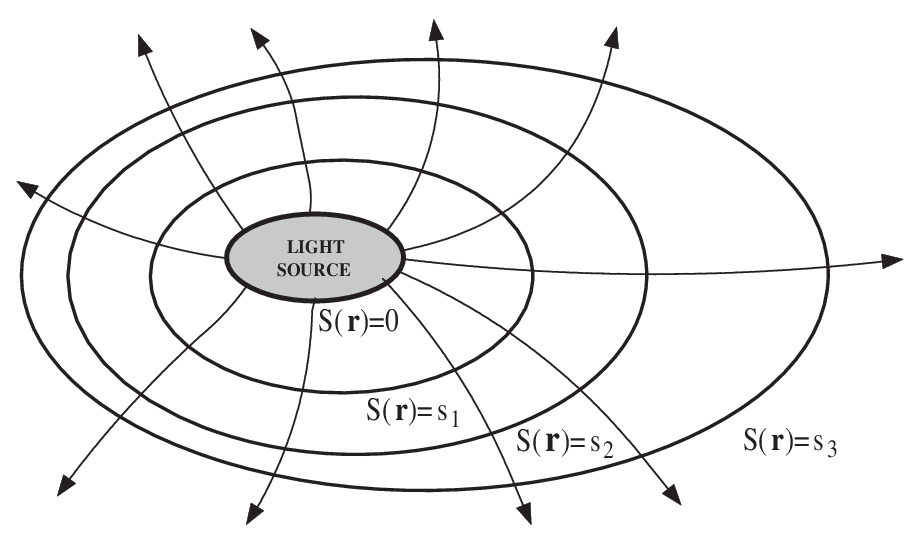}
\end{center}
\caption{Wavefronts and rays emanating from a light source.}
\label{fig:rays}
\end{figure}

It follows from Maxwell's equations that the function $S(\br)$ satisfies the following non-linear partial differential equation, known as the \emph{eikonal equation}:
\begin{equation}
(\frac{\partial S}{\partial x})^2+ (\frac{\partial S}{\partial y})^2=n^2 \Leftrightarrow (\nabla S)^2=n^2 \Leftrightarrow |\nabla S |=n.
\label{eq2}
\end{equation}
In the above, $n=n(\br) \geq 1$ is the \emph{refractive index}, a unit-less scalar quantity that captures the properties of the medium at location $\br$. The vector $\nabla S \triangleq \frac{\partial S}{\partial x} \hat{\textbf{x}}+ \frac{\partial S}{\partial y} \hat{\textbf{y}}$ is the gradient of $S$. Its direction gives the direction along which $S$ increases the fastest, and its absolute value $|\nabla S|$ gives the rate of this increase. To understand the meaning of (\ref{eq2}), note that at a point $\br$, $|\nabla S| =\frac{dS}{ds}=\frac{dt \times c}{ds}$, where $ds$ is an incremental distance along the direction of maximum increase of $S$, $dS$ is its increase along $ds$, and $dt$ is the time it takes the light to cover distance $ds$. Applying (\ref{eq2}), it follows that $\frac{ds}{dt}=\frac{c} {n(\br)}$, i.e., equation (\ref{eq2}) specifies that at the location $\br$ light travels with speed $\frac{c}{n(\br)}$. The initial condition needed for solving (\ref{eq2}) is that $S(\br)=0$ along the set of points $\Omega$ representing the source. Having this initial condition, the eikonal equation can then be solved for all points $\br$ outside the source.

We briefly note that the eikonal equation is of interest not only in the context of Geometrical Optics, but can be used to model a large number of disparate phenomena in disciplines such as seismology, fluid mechanics, and material science~\cite{sethian1}. As an off-the-wall example, the eikonal equation can be used to predict the way a large group of termites will eat up wood, assuming that they are initially placed in a domain $\Omega$, and are capable of eating up the wood at point $\br$ with speed $\frac{c}{n(\br)}$, where $c$ is a nominal speed, and $n(\br)>0$ is the wood hardness. 

Having defined the eikonal function and the related notion of wavefronts, we can define \emph{light rays} as the set of curves that originate at the light source and move away from it, while keeping constantly orthogonal to wavefronts. An example is shown in Fig.~\ref{fig:rays}. Intuitively, light rays are the paths that light takes as it emanates away from the source.

By applying the eikonal equation (\ref{eq2}), it can easily be seen~\cite{born1} that light rays satisfy the following non-linear second-order differential equation:
\begin{equation}
\frac{d}{ds}(n \frac{d\br}{ds})=\nabla n,
\label{eq1}
\end{equation}
where $s$ is the length of the ray measured from a fixed point to position $\br$, $\frac{d\br}{ds}$ is the direction of the ray at  $\br$, $n(\br)$ is the refractive index at $\br$, and $\nabla n = \frac{\partial n}{\partial x} \hat{\textbf{x}} +\frac{\partial n}{\partial y} \hat{\textbf{y}}$ is its gradient. Note that the gradient $\nabla n$ points toward the direction where the refractive index increases. Therefore, (\ref{eq1}) expresses the fact that light rays tend to bend toward the direction where the medium is optically denser, i.e., the index of refraction has a higher value. Equation (\ref{eq1}) holds wherever the refractive index is a continuous function of the position. Locations with discontinuities can be handled by applying (\ref{eq2}) together with limiting arguments, leading to Snell's laws of reflection and refraction \cite{born1}. 

Next, let the \emph{optical length} $[AB]_C$ of a curve $C$ connecting two points $A$ and $B$ be the line integral
\begin{equation}
[AB]_C=\int_A^B n(\br)\,ds 
\label{eq3}
\end{equation}
along the curve $C$. Since the speed with which light moves at a point $\br$ is $u= \frac{c}{n(\br)}$, the time needed by light to go from $A$ to $B$ or vice versa along the curve $C$ equals 
\begin{equation*}
\int_A^B\,dt=\int_A^B \frac{1}{u}\,ds=\frac{1}{c} \int_A^B n(\br) \,ds=\frac{[AB]_C}{c}.
\end{equation*}
Therefore, the optical length of a curve equals the time it would take light to travel across the curve $C$, multiplied by the speed of light in free space $c$.

\begin{figure}
\begin{center}
\includegraphics[scale=0.5]{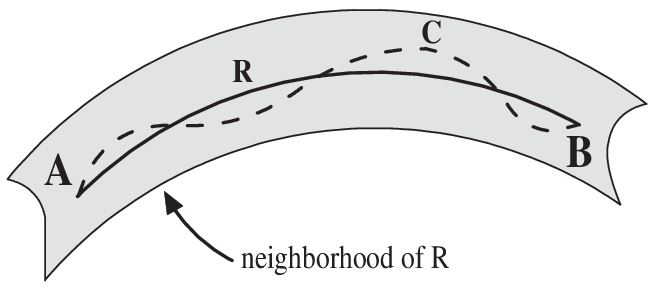}
\end{center}
\caption{Illustration of the principle of Fermat: Light traveling from $A$ to $B$ along any small perturbation $C$ of a light ray $R$ would necessarily take more time.}
\label{fig:fermat}
\end{figure}

Now let the source of light be at a point $A$, and let $R$ be a curve starting at $A$ and ending at some other point $B$, as shown in Fig.~\ref{fig:fermat}. The \emph{principle of Fermat} states that the curve $R$ will be a ray of light if and only if its optical length is smaller than the optical length of any other curve $C$ that passes through $A$ and $B$ and exists within a sufficiently small neighborhood of the ray $R$. In other words, if we introduce a sufficiently small perturbation in a ray $R$, resulting in a new curve $C$, then necessarily the new curve will have a larger optical length. As the optical length is proportional to the time needed to go from $A$ to $B$, the principle of Fermat asserts that light takes the fastest route. 

\section{Macroscopic Network Model} \label{MacroscopicModel}

Our network consists of nodes placed over a 2-dimensional domain $\s{D}$. We assume that there are so many nodes, that the typical distance between nearest neighbors is incremental with respect to the dimensions of $\s{D}$.

We associate with the movement of packets over the network a consumption of resources which is cumulative with the distance over which the packet is transported. The resources could be the bandwidth used, the energy wasted, the time needed for the transport, or some other composite of the above. However, in this section, we refrain from further specifying the nature of the resources, and we keep the discussion abstract. We model the use of resources in terms of a scalar \emph{cost function} $c(\br)$, which is unit-less, and is defined as follows: Let $\epsilon$ be the incremental length of a small line segment centered at $\br$. Let $dc(\epsilon,\br)$ be the incremental cost associated with transporting a packet along this incremental line segment. Let $c_{\mathrm{nominal}}\times \epsilon$ be the cost incured by transporting the packet in some given \emph{nominal network} by a distance $\epsilon$. The cost function is defined as follows:
\begin{equation}
c(\br)=\lim_{\epsilon \rightarrow 0} \frac{dc(\epsilon,\br)}{c_{\mathrm{nominal}}\times \epsilon}.
\label{eq6}
\end{equation}
For example, a cost function $c(\br)=2$ means that it is twice as costly to move packets along the location $\br$ of our network with respect to the nominal network. 

Initially, this definition of the cost function may seem counterintuitive. Indeed, it would make more sense to simply define the cost function as the incremental cost over the incremental distance, without dividing by $c_{\mathrm{nominal}}$. The advantages of our approach are two-fold: firstly, in this manner the cost function is unit-less, irrespective of the precise nature of the cost. This makes the definition more general. Secondly, we will soon show an analogy of our setting with Optics, where the nominal network is mapped to free space, and the cost function is mapped to the refractive index. As the refractive index is unit-less, making the cost function also unit-less makes the analogy tighter. 

Using the definition of the cost function (\ref{eq6}), it follows that the cost to move a packet from point $A$ to point $B$, along a curve $C$ of non-incremental length, is given by the line integral of the cost function along $C$:
\begin{equation}
[A B]_C=\int_A^B c(\br) \,d\br.
\label{eq4}
\end{equation}
The cost is measured in meters, and represents the distance that will be covered in the nominal network, using the same volume of resources. 

We apply the above concepts in two relevant scenarios:
\begin{itemize}
\item \textbf{Simple Routing Problem:} Let $A$ be a node in the network. Given any other node $B$ in the network, what is the minimum cost route connecting the two, and what is the associated minimum cost?
\item \textbf{Distributed Routing Problem:}  Let $\Omega \subset \s{D}$ be a sub-domain of  $\s{D}$ containing a set of nodes in the network. Given any other node $B$ in the network, what is the minimum cost route connecting $B$ with $\Omega$, and what is the associated minimum cost?
\end{itemize}

Note that the second question is fundamentally different from the first question, because in the first case we want to find a route connecting two known locations, but in the second case the one end of the route is not known. The second case is relevant where we need to find a minimum cost route, but we have some freedom in selecting one of its end points. For example, the set $\Omega$ might be a collection of points corresponding to the cluster heads of a wireless ad hoc network. In such a case, the rest of the nodes will want to communicate with that of the cluster heads which incurs the minimum cost. As another example, the set $\Omega$ might represent a distributed set of sinks that are placed in a wireless sensor network. Individual sensors will want to relay the information they have collected to the nearest sink. 

\subsection{Analogy with Geometrical Optics}

Both problems involve the minimization of the integral (\ref{eq4}) over all possible curves $C$, subject to constraints on the start and end of the curves, and can be solved with Calculus of Variations techniques. Luckily for us, both of these two problems have been solved and extensively investigated in the context of Optics. 

In more detail, optical media are described in terms of the refractive index $n(\br)$: in the time it takes light to cover a distance of $\epsilon$ in the medium, light can cover a distance of $\epsilon n(\br)$ in free space. Networks, on the other hand, are described in terms of the cost function: with the cost it takes the network to transport the packet across a distance $\epsilon$, the packet could be transported across a distance $\epsilon c(\br)$ in the nominal network. Therefore, going from Optics to Networking, the optical medium becomes the network, the refractive index becomes the cost function, the optical length of a curve, given by (\ref{eq3}), becomes the cost to transport a packet along that curve, given by (\ref{eq4}), and free space becomes the nominal network. 

We can then solve the Simple Routing Problem by framing it into the context of Optics: Finding the curve with the minimum cost is equivalent to finding the curve with the minimum optical length in an optical medium with refractive index $n(\br)=c(\br)$ and a source of light at point $A$. However, as already discussed, Fermat's principle specifies that this curve is actually a ray of light that starts from $A$ and goes to $B$! Therefore, the optimal route satisfies the differential equation (\ref{eq1}), if we substitute $n(\br)$ with $c(\br)$, together with the initial conditions that it must start at $A$ and end at $B$.

We can also solve the Distributed Routing Problem in a similar manner. In particular, we need the minimum costs, and the routes that achieve them, between the region $\Omega$ and all other points in the network. In the context of Optics, we need to find the minimum optical lengths, and their corresponding rays, between a distributed light source occupying $\Omega$ and all other points $B$ in an optical medium whose refractive index is $n(\br)=c(\br)$. By definition, these minimum optical lengths are given by the eikonal function, which in turn can be found by solving the eikonal equation (\ref{eq2}). Knowing the eikonal function, we can costruct the wavefronts and from these the rays that connect the source with any part of the network.

\section{Microscopic Network Models} \label{MicroscopicModel}

In the previous section we developed a macroscopic network model, built around the notion of an abstract cost function, which we left unspecified. In this section, we develop two microscopic models, specifying, among others, the resources consumed by each transmission, the model for node placement, and the rules of forwarding packets. We then calculate the cost function that corresponds to each of these microscopic models. 

\subsection{Bandwidth Limited Networks}

We consider first the case in which the most scarce resource is bandwidth and therefore the objective of the routing protocol is to minimize bandwidth consumption.

\textbf{Node Deployment:} We assume that the nodes are placed randomly, according to a spatial Poisson distribution \cite{ross1, franceschetti1} with density $\lambda(\br)$ measured in nodes per square meter and defined by the following limit:
\begin{equation*}
\lambda(\br)=\lim_{ \epsilon \rightarrow 0} \frac{E[N(\s{B}(\br,\epsilon))]}{\pi \epsilon^2},
\end{equation*}
where $\s{B}(\br,\epsilon)$ is the ball centered at $\br$ with radius $\epsilon$, $N(\s{B}(\br,\epsilon))$ the number of nodes within that ball, and $E[\cdot]$ denotes the expectation of a random variable.

\textbf{Transmission cost:} We assume that the nodes communicate with each other through a common wireless channel of finite bandwidth. When a node $T$ transmits to a node $R$, nodes in the vicinity must refrain from transmitting, in order to avoid collisions. 

In this setting, it is reasonable to assume that each transmission over a distance $d$ occupies an area proportional to $d^2$. An intuitive justification for this is as follows: If a node $T$ transmits to a node $R$ over some distance $d$, then it is necessary for all other potential transmitters that are in a neighborhood of the receiver to refrain from transmitting. Intuitively, the radius $r$ of this neighborhood is proportional to the distance $d$. Indeed, the larger $d$ is, the larger $r$ should be to ensure a minimum acceptable Signal to Interference Ratio (SIR) at the receiver. Therefore, the area of the neighborhood is $k d^2$, where $k$ is a unit-less constant that depends on the sensitivity of the receiver, the rate used, etc. A rigorous derivation of the consumed area for a particular reception model exists in \cite{gupta1}, but is not replicated here for lack of space. 

Clearly, as the areas occupied by communicating pairs in a network increase in size, the number of simultaneous transmissions decreases, as the total occupied area can not exceed the total area of the network. In turn, this drives the aggregate throughput down\footnote{We note that the proof for the upper bound of the capacity of a wireless network in \cite{gupta1} was based on exactly this idea.}. It would be reasonable for network designers to keep the size of the occupied areas as small as possible, in order to maximize throughput, and so it is reasonable to define the \emph{transmission cost} of a transmission over a distance $d$ to be 
\begin{equation}
c_t=d^2.
\label{eq7}
\end{equation}

\textbf{Forwarding Rule:} To complete our microscopic model we must also specify the rule according to which nodes forward packets toward their destination. Microscopic parts of trajectories can be viewed as straight lines, therefore it suffices to have a rule for forwarding packets along a given direction. With no loss of generality, let this direction be that of the positive $x$-axis. 

\begin{figure}
\begin{center}
\includegraphics[scale=0.52]{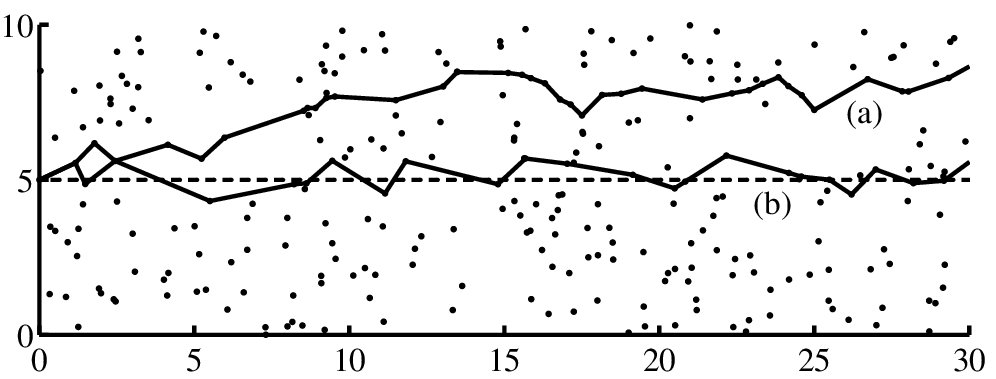}
\end{center}
\caption{ Packet forwarding along a straight trajectory according to two different forwarding rules: \textbf{Rule A:} (route (a)) The cost-progress ratio $\frac{c_t}{x}$ must be minimized. \textbf{Rule B:} (route (b)) The cost-progress ratio  $\frac{c_t}{x}$ must be minimized, but packets must stay also close to the trajectory.} 
\label{fig:rules}
\end{figure}

Given that we have associated each transmission over a distance $d$ with a cost $c_t=d^2$, a reasonable first choice for the node is to forward the packet to that of its neighbors closer to the destination (along the trajectory) for which the ratio 
\begin{equation*}
r \triangleq \frac{c_t}{x_2-x_1}=\frac{d^2}{x_2-x_1}
\end{equation*} 
is minimized. In the above, $d$ is the distance between the node and its neighbor, $x_1$ is the $x$-coordinate of the node, and $x_2$ is the $x$-coordinate of the neighbor. The difference $x_2-x_1>0$ represents the progress toward the destination at the end of the trajectory. Therefore, we would like to strike a balance between incuring a small transmission cost, but also achieving a substantial progress toward the destination. We will refer to this forwarding rule as Rule A. In Fig.~\ref{fig:rules} we plot a sample route (route (a)) that would result from this forwarding rule. 

As shown in the figure, a problem with Rule A is that it may cause the packet to deviate significantly from its trajectory, over the course of many hops. (In fact, it is intuitively clear that the deviation from the trajectory can be modeled by a random walk where the steps take continuous values.) Therefore, we are motivated to consider the following modification: if the node forwarding the packet is on the left (right) of the trajectory and at some distance $v$ from it, then the next node to receive the packet can not be also on the left (right) of the trajectory and at some distance $v'>v$ from it. The rest of the nodes are acceptable as next hops, and the one minimizing the cost-progress ratio should be chosen. An example route appears in Fig.~\ref{fig:rules} (route (b)). As the figure shows, the new rule leads to the packets following the trajectory much more closely. 

A disadvantage of Rule B with respect to Rule A is that we are excluding from our consideration neighbors with a very small associated cost-progress ratio, but who happen to increase the deviation from the ideal trajectory even very modestly. The observant reader will notice a few such examples in Fig.~\ref{fig:rules}. Obviously, there is significant room for improving both Rule A and Rule B. As the question of choosing the next hop within the context of TBF and geographic routing has already attracted significant research interest (see for example \cite{lee3} and the references therein), we do not pursue this issue further. In the following, we will assume that Rule B is used. 

\textbf{Forwarding Rule Performance:} Next, we evaluate the performance of Rule B in terms of the expected progress and the expected cost per hop. With no loss of generality, we assume that the node having the packet is placed at the origin, and that the trajectory is along the positive $x$-axis, and described by the equation $y=\mathrm{const}>0$. The progress is then simply the $x$-coordinate of the next node, and in order to have a positive progress toward the destination, the next hop must have a positive $x$-coordinate. In addition, in order to satisfy Rule B, it must also have a positive $y$-coordinate. Therefore, only nodes in the positive quadrant will be considered for forwarding the packet. 

As the placement of nodes is random, the location of the next node will also be random, and so its associated progress, $y$-coordinate, distance from the origin, cost, and cost-progress ratio are random variables, which we denote by $X$, $Y$, $D$, $C$, and $R$ respectively. 

\begin{figure}
\begin{center}
\includegraphics[scale=0.49]{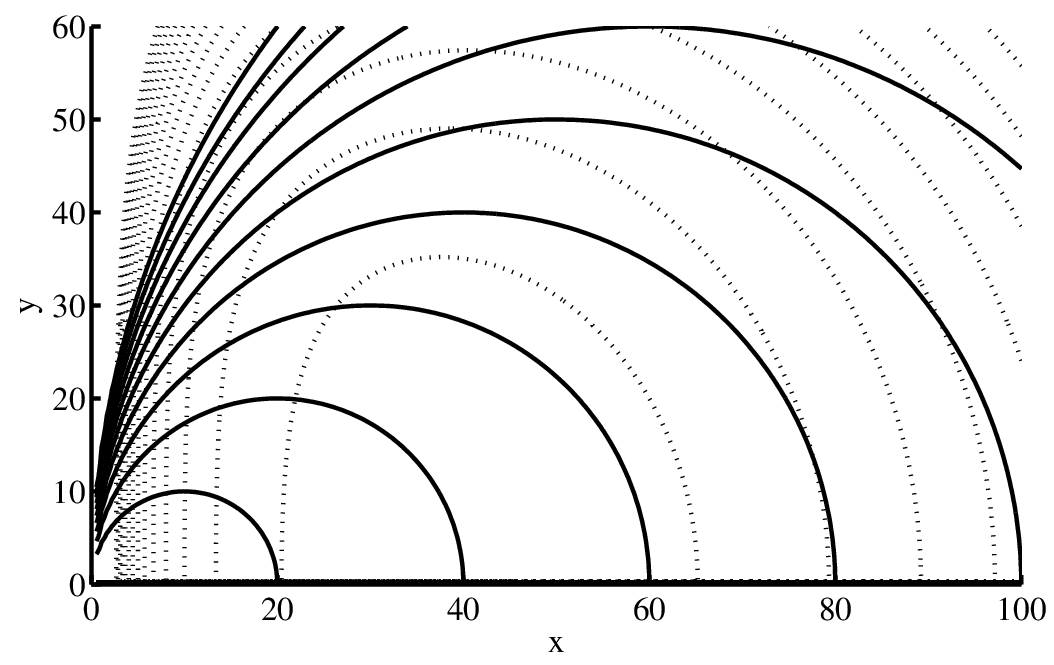}
\end{center}
\caption{Loci of constant cost-progress ratio in the bandwidth limited case (continuous lines) and energy-limited case (dashed lines).}
\label{fig:loci1}
\end{figure}

The first step is to find the loci of constant cost-progress ratio. With straightforward algebra, it follows that:
\begin{equation*}
\frac{d^2}{x}=r \Leftrightarrow (x-\frac{r}{2})^2+y^2=(\frac{r}{2})^2.
\end{equation*}
Therefore, as shown in Fig.~\ref{fig:loci1}, the loci of constant ratio are semi-circles centered at point $(\frac{r}{2},0)$ with radius $\frac{r}{2}$. Note that as $r \rightarrow 0$, the corresponding locus implodes toward the origin. Therefore, to minimize the cost-progress ratio, the node would like to avoid transmitting to nodes further away. This is consistent with the long-known fact that many short hops are better than a few long ones, if the bandwidth is limited~\cite{gupta1,kleinrock87}.

Knowing the shape of the loci, we can calculate the distribution of the ratio $R$. The ratio will be larger than $r$ if and only if there is no node within a semi-circle of radius $\frac{r}{2}$. As the nodes follow a Poisson spatial distribution with density $\lambda(\br)$, this happens with probability $\exp[-\frac{\lambda(\br)\pi r^2}{8}]$ . Therefore, 
\begin{equation*}
F_R(r)=1-e^{-\frac{\lambda(\br)\pi r^2}{8}} \Leftrightarrow
f_R(r)=\frac{r  \lambda(\br)\pi}{4}e^{-\frac{\lambda(\br)\pi r^2}{8}}.
\end{equation*}
So the ratio $R$ follows the Rayleigh distribution, with expectation $E[R]= \sqrt{\frac{2}{\lambda(\br)}}$.

Next, we calculate the expectation $E[X]$ of the progress, using conditioning on $R$. In particular, let us assume that $R=r$. It then follows that the next node is on the circumference of the semi-circle centered at $(\frac{r}{2},0)$ and with radius $\frac{r}{2}$. By symmetry, it follows that the conditional expectation of its $x$-coordinate will be $E[X|R=r]=\frac{r}{2}$. Therefore:
\begin{eqnarray*}
E[X] &=& \int_0^\infty E[X|R=r]f_R(r)\,dr \\
&=& \int_0^\infty \frac{r^2 \lambda(\br) \pi}{8}  \exp[-\frac{\lambda(\br)\pi r^2}{8}]\,dr = \frac{1}{\sqrt{2\lambda(\br)}}.
\end{eqnarray*}

\begin{figure}
\begin{center}
\includegraphics[scale=0.55]{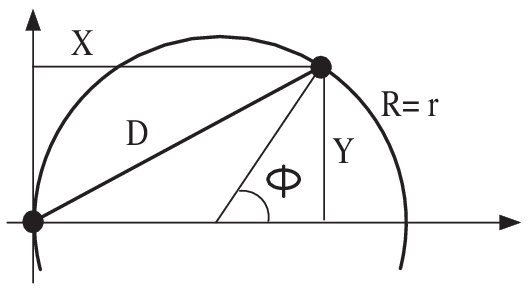}
\end{center}
\caption{Random variables associated with the next hop.}
\label{fig:hop_notation}
\end{figure}

Finally, we calculate the expectation $E[C]= E[D^2]$ of the cost, again by conditioning on the value of the ratio $R$. For this, we need to calculate $E[D^2|R=r]$. We note that $D^2=\frac{R^2}{2}(1+\cos \Phi)$ where, as shown in Fig.~\ref{fig:hop_notation}, $\Phi$ is the angle formed by the positive $x$-axis and the line interval connecting the points $(X,Y)$ and $(\frac{r}{2},0)$. Given that $R=r$, by symmetry considerations the angle $\Phi$ is uniformly distributed in $(0,\pi)$. It follows that 
\begin{equation*}
E[D^2|R=r]=\frac{1}{\pi} \int_0^{\pi} \frac{r^2}{2}(1+\cos \phi)\,d\phi=\frac{r^2}{2}.
\end{equation*}
Therefore, 
\begin{eqnarray*}
E[D^2]&=&\int_0^\infty E[D^2|R=r]f_R(r)\,dr \\
&=& \int_0^\infty \frac{r^3 \pi \lambda(\br)}{8}\exp[-\frac{\lambda(\br)\pi r^2}{8}]\,dr 
= \frac{4}{\pi \lambda(\br)}.
\end{eqnarray*}

\textbf{Cost Function:} Having determined the performance of the forwarding rule, we are now ready to specify the cost function for this microscopic model. For this, let us assume that a packet following the forwarding rule gets transported across an incremental distance of $\epsilon$. By the massively dense assumption, this distance is covered with a large number of hops. To calculate the associated incremental cost, we note that the sequence of progresses $\{X_i\}$ together with their associated costs $\{C_i\}$ form a reward renewal process, where $C_i$ is the reward of taking a step of size $X_i$. By the strong law of large numbers for reward renewal processes \cite{ross1}, it follows that $\frac{\sum_{i=1}^n C_i}{x} \rightarrow \frac{E[D^2]}{E[X]}$ as $x,n \rightarrow \infty$. By the massively dense assumption, the limit is achieved for $x=\epsilon$, therefore the incremental cost $\sum_{i=1}^n C_i=\frac{E[C]}{E[X]}\epsilon = \frac{4\sqrt{2}}{\pi} \frac{1} {\sqrt{\lambda(\br)}} \epsilon$. To keep the expressions simple, let us assume that $c_{\mathrm{nominal}}=\frac{4\sqrt{2}}{\pi}$. It follows that
\begin{equation*}
c(\br)=\frac{1}{\sqrt{\lambda(\br)}}.
\end{equation*}

\begin{figure}
\begin{center}
\includegraphics[scale=0.49]{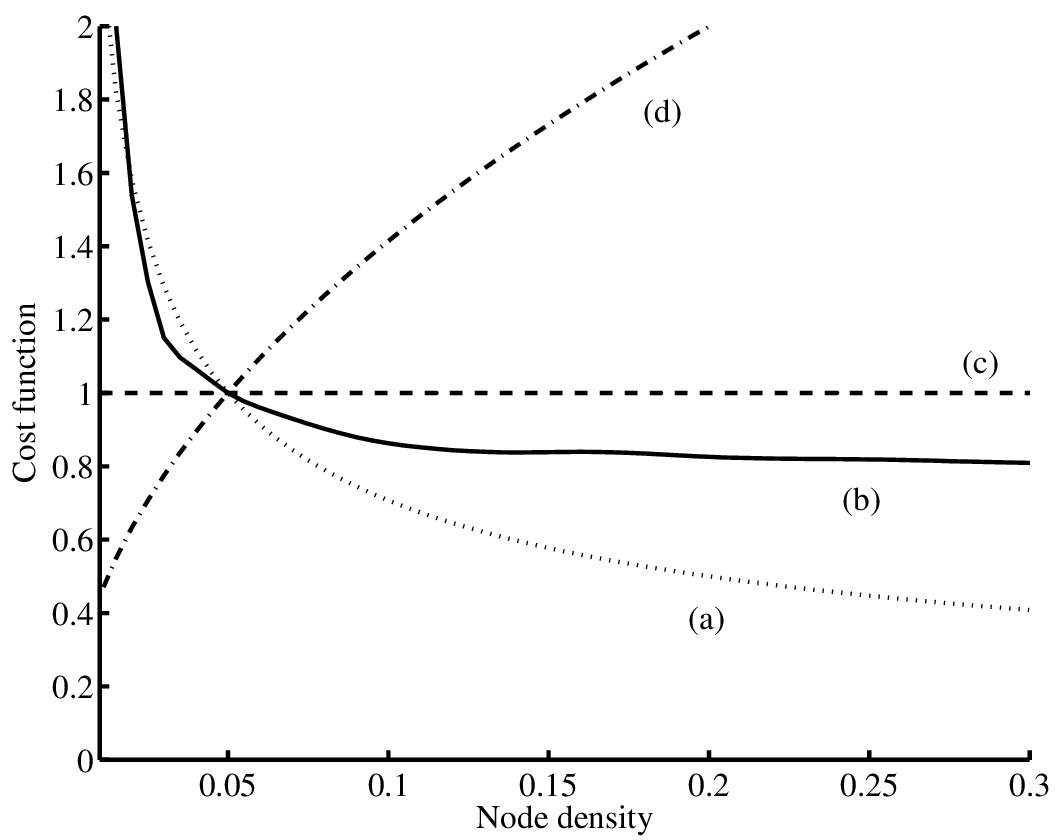}
\end{center}
\caption{The cost function versus the density for four microscopic models: \textbf{(a)} Bandwidth Limited, \textbf{(b)} Energy Limited, \textbf{(c)} Nominal network, \textbf{(d)} Minimum hop model of \cite{jacquet1}. Costs are normalized so that they all intersect at the point $(0.05,1)$.}
\label{fig:costs}
\end{figure}

The result is intuitive. Indeed, let us assume that the density is increased by a factor of $4$. Then the expected length of a transmission decreases by a factor of $2$, hence the area occupied by a transmission decreases by a factor of $4$. However, now we need twice as many transmissions to cover the same distance. Therefore, if we increase the node density by a factor of $4$, the overall area consumed decreases by a factor of $2$. This back-of-the-envelope calculation is consistent with the cost function we arrived at.

\subsection{Energy Limited Networks}

Let us now consider a network of nodes communicating over a common wireless channel with very large, practically infinite bandwidth. This is the case, for example, with nodes using Ultra-Wide-Band transceivers. In such a case, each transmitter can use its own dedicated bandwidth, and communication is not hampered by interference~\cite{negi1}.

On the other hand, we assume that nodes have a limited energy supply. Therefore, we would like from the network to transport data in an energy efficient manner. We assume that nodes communicate using a fixed data rate, and in order to reach a receiver at a distance $d$, the transmission incurs an energy cost
\begin{equation}
C(d)=ad^b+c.
\label{eq8}
\end{equation}
This is the standard model used in the literature~\cite{chang1, ephremides3}. The constant term $c$ captures the energy consumed in the electronics of the transmitter and receiver, which is independent of the transmission distance. The term $ad^b$ captures the energy consumed by the transmitter power amplifier, and which depends on the distance. The exponent $b$ typically obtains its value in the range $(2,6)$, depending on the environment~\cite{rappaport}. 

Under this energy consumption model, transporting a packet across the network using a large number of very short hops incurs a very large aggregate energy cost, due to the constant term $c$. Likewise, transporting a packet with a few long hops also incurs a large energy cost, due to the term $ad^b$ which typically increases with distance much faster than linearly. In fact, it is intuitively clear that if the nodes could decide on the hop distance, they would select the optimal value $d_{\mathrm{opt}}$ for which the ratio $\frac{ad^b+c}{d}$ is minimized. Straightforward calculation shows that $d_{\mathrm{opt}}=\left[\frac{c}{(b-1)a} \right]^\frac{1}{b}$.

The forwarding Rule B used in the bandwidth limited model continues to make sense in this case, with the modification that the transmission cost is given by (\ref{eq8}). As in the previous case, we would like to determine the expected cost and progress per hop incured by this forwarding rule, and use them to establish the cost function. 

A complication arises from the fact that the constant cost-progress loci $\frac{ad^b+c}{x}=r$ are no longer circles. In Fig.~\ref{fig:loci1} we plot the loci for a typical choice of parameter set. As shown in the figure, the loci are oval shaped, roughly centered at the point $(d_{\mathrm{opt}},0)$. As a result, the method of the previous section for finding the expected cost and progress can not be replicated. Nevertheless, we can easily calculate the expected cost and progress numerically, by Monte Carlo simulation. In Fig.\ref{fig:costs} we plot the resulting cost function versus the node density.

As in the previous case, the resulting cost function is a decreasing function of the node density. Contrary to the previous case, however, as the node density goes to infinity, the cost function converges to a strictly positive limit. This is expected: After the node density achieves a sufficiently high value, each node is guaranteed to find a node further down the trajectory and at a distance very close at the optimal $d_{\mathrm{opt}}$. After that point, increasing the density further does not affect the cost function, as the optimal point of operation has already been achieved. 

\subsection{Other Models}

Our bandwidth limited and energy limited models are only two of the many possibilities. Using the same methodology, many other choices are possible. We briefly mention that in \cite{jacquet1} the cost function $c(\br)=\sqrt{\lambda(\br)}$ was implicitly used. As discussed, that cost function corresponds to the case where we want to minimize the number of hops, while restraining the communication between nearest neighbors. An alternative cost function was also introduced in \cite{catanuto1}, under a formulation less developed that our own. That cost corresponded in the case where we want to minimize the energy needed for the transport of a packet, but all nodes are required to transmit with fixed power, i.e. the parameter $a$ of (\ref{eq8}) is set to $a=0$. The authors of \cite{baccelli2} propose, independently from our line of work, a cost function that captures the energy needed for routing in a large-scale static interference field. 

Alternatively, the cost function could be specified in an adaptive manner, depending on the fluctuating needs and capabilities of the network. For example, we could perform adaptive routing by using a cost function that captures the congestion at the various parts of the network. Therefore, congested areas could select a high value for their cost function, and so discourage the use of routes through them. Areas of the network with spare resources could select low values for their cost function, thereby inviting routes through them. This is a potential subject for future work.

\section{Simulation Results} \label{Simulations}

In this section, we present a few examples of optimal routes, under the different microscopic models we have considered. Due to space constraints, this section is only intended to provide a proof-of-concept and illuminate the theory, rather than exhaustively investigate the performance of protocols based on the Optics analogy.

\begin{figure}
\begin{center}
\includegraphics[scale=0.48]{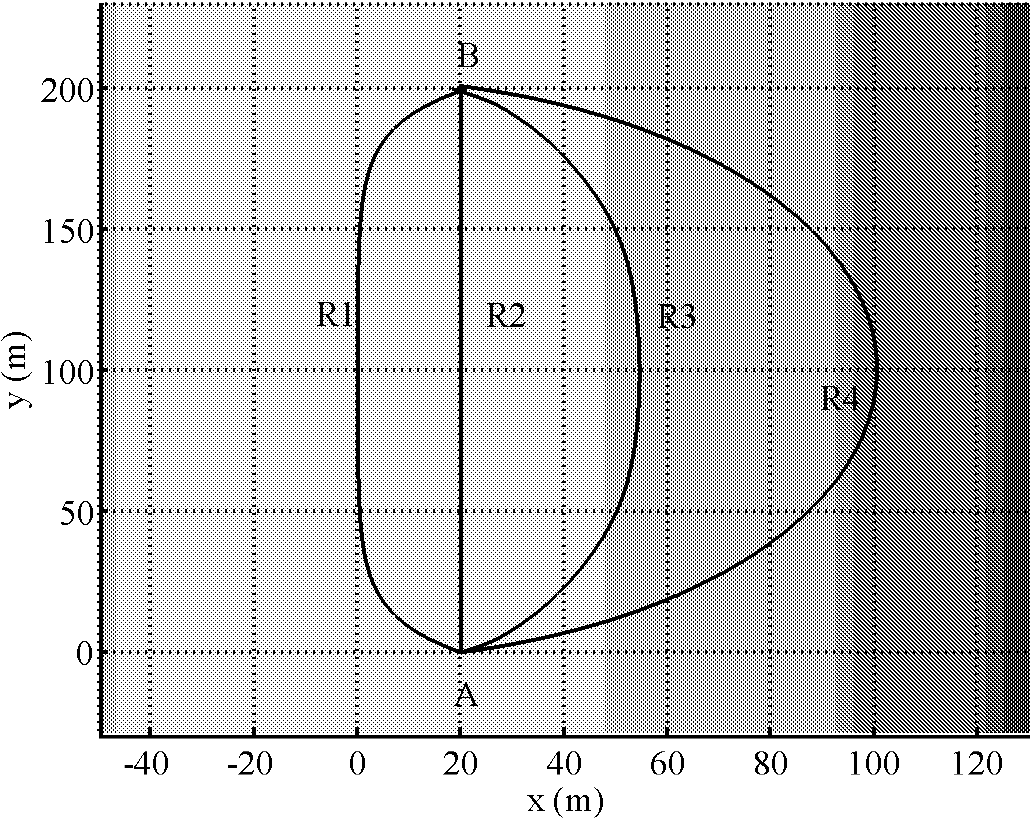}
\end{center}
\caption{Optimal routes under four microscopic network models: \textbf{(R1)} The model of \cite{jacquet1}, \textbf{(R2)} Constant cost function model \textbf{(R3)} Energy limited model, \textbf{(R4)} Bandwidth limited model.}
\label{fig:routes}
\end{figure}

We consider a network in which the node density $\lambda(\br)=3 \times 10^{-5}x^2 +0.01$, i.e., there is no dependence of the density on the $y$-coordinate. In Fig.~\ref{fig:routes} we plot the optimal routes connecting the two points $A=(20,0)$ and $B=(20,200)$ for the bandwidth-limited network model (Route R4), the energy limited model (route R3), the model of \cite{jacquet1} (Route R1), and a network in which the cost function is constant and independent of the node density. In the last case, the optimal route is simply the straight line connecting the points $A$ and $B$. In the figure, areas with high node density appear darker.

As expected, under the cost model of \cite{jacquet1}, the optimal route goes through the region of the network with very low density. Also as expected, in both the energy-limited and the bandwidth-limited cases the optimal routes  (R4 and R3 respectively) are through the part of the network with high concentration. However, R3 does not deviate as much as R4. The reason is that, in the bandwidth limited case, the cost function $c$ decreases with the density for arbitrarily high values of the density $\lambda$ as $c=\frac{1}{\sqrt{\lambda}}$ (see Fig.~\ref{fig:costs}). For the energy limited case, however, after the density reaches some value, a further increase does not reduce the cost. Therefore, deviating through regions of very high density is not beneficial and the optimal routes change accordingly.

For the bandwidth limited case, the aggregate cost of the straight line R2 is $80\%$ more than the aggregate cost of the optimal route R4, even though R4 is significantly longer. Similar gains exist with respect to the straight line for the other cases. 

\begin{figure}
\begin{center}
\includegraphics[scale=0.50]{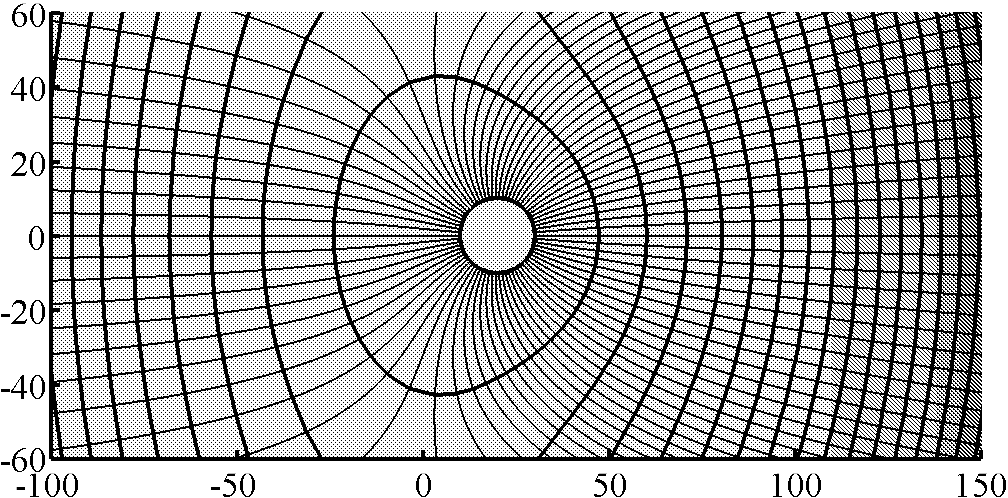}
\textbf{(a)}
\includegraphics[scale=0.50]{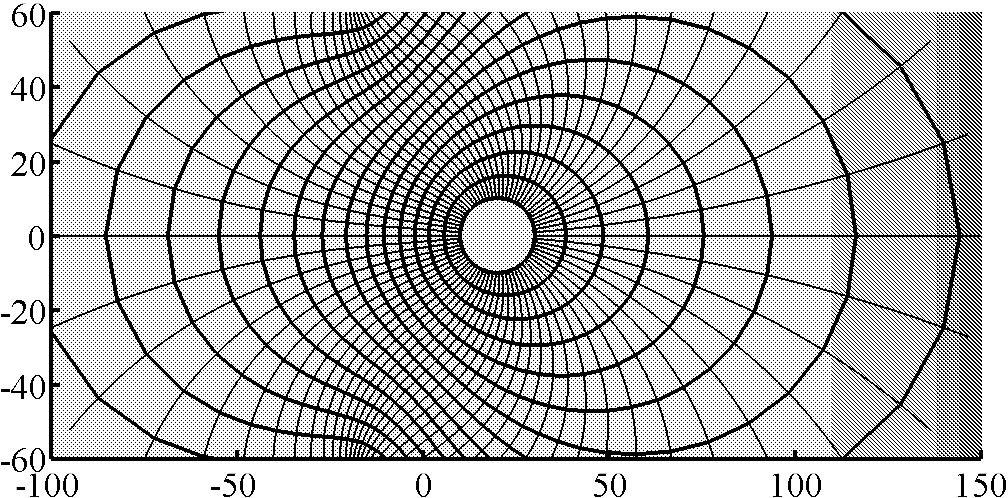}
\textbf{(b)}
\end{center}
\caption{Wavefronts and rays in a case of distributed routing, and for three microscopic network models: \textbf{(a)} The model of \cite{jacquet1}, \textbf{(b)} Bandwidth limited model.}
\label{fig:eikonals}
\end{figure}

In Fig.~\ref{fig:eikonals}, we plot wavefronts and rays in an example of distributed routing and for two microscopic models: the bandwidth limited, and the model of \cite{jacquet1}. The region $\Omega$ is a circle centered at the point $(20,0)$, with radius $10$.  As expected, under the bandwidth limited model, regions of low node density incur high costs, and for this reason routes bend so that they pass this region as fast as possible. The opposite happens for the model of \cite{jacquet1}.

\begin{figure}
\begin{center}
\includegraphics[scale=0.56]{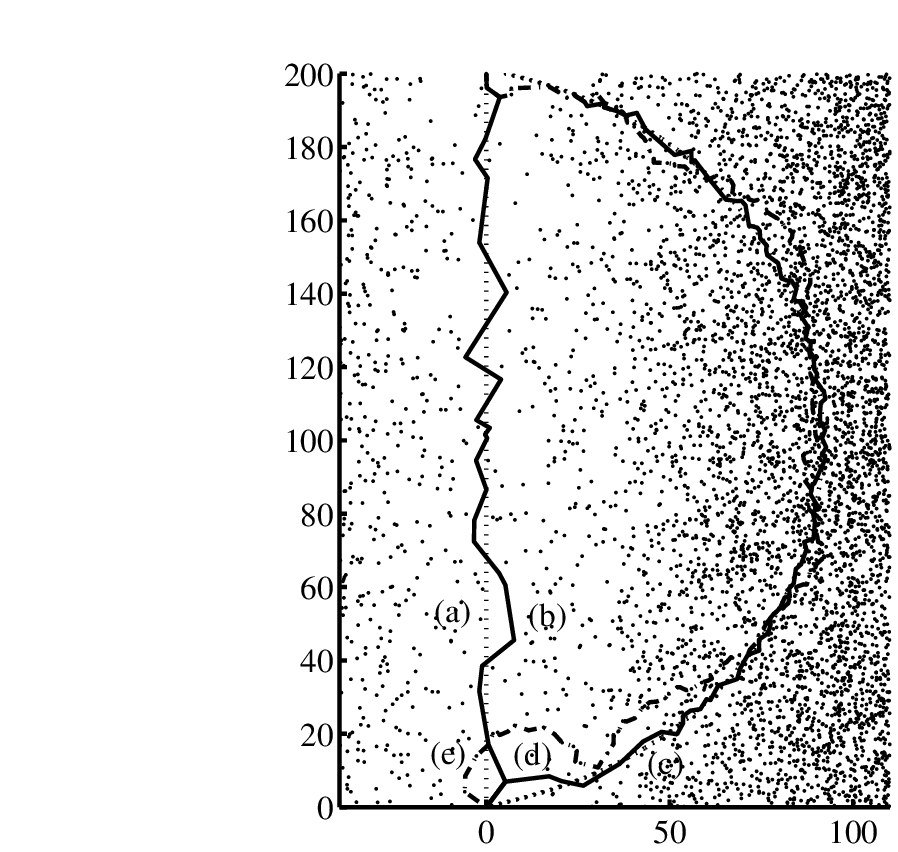}
\end{center}
\caption{A realization of a massively dense network, with roughly $5000$ nodes, and with a source placed at $(0,0)$ and a destination placed at $(0,200)$. \textbf{(a)} Straight line connecting the source and destination. \textbf{(b)} Trajectory based forwarding along the straight line. \textbf{(c)} Curve determined by the Optics analogy \textbf{(d)} Trajectory based forwarding on curve (c). \textbf{(e)} Optimal route determined by Bellman-Ford algorithm.}
\label{fig:trajectories}
\end{figure}

Finally, in Fig.~\ref{fig:trajectories} we consider a particular realization of a network with node density $\lambda(\br)=0.5 \times 10^{-4}x^2+0.025$, and we compare \emph{(i)} the route that results from performing TBF according to Rule B on the route determined by the Optics analogy, with \emph{(ii)} the route that results from performing TBF on the straight line connecting the source and the destination, and \emph{(iii)}  the optimal route, determined by the Bellman-Ford algorithm. We note that the optimal route follows closely the route determined by Optics. That should be expected, as our route provably becomes optimal in the limit of a large number of nodes. The costs of the three routes are $1305$, $2750$, and $944$ respectively. Therefore, optical routing needs less than half the resources needed when routing along the shortest path, but $40\%$ more resources with respect with the optimal route. We note, however, that calculating the optimal route becomes more and more impractical as the number of nodes increases. 

\section{Conclusions} \label{Conclusions}
In this work, we investigate the problem of optimally selecting routes in massively dense networks. We use the observation of \cite{jacquet1} that the optimal routing problem can be cast as a problem in Optics, significantly extending the theory and methodology of that work. Although in practice it may not be possible to implement our routing algorithms in networks with modest size and computing capabilities, still, knowing the limiting optimal routes allows a meaningful evaluation of simpler, suboptimal algorithms, and provides useful insights for the development of appropriate routing rules.  

Due to lack of space, a number of important issues remain unaddressed. First of all, although Geometrical Optics describes optimal routes in a very elegant manner, it does not provide the tools for a source-destination pair to discover, in a distributed manner, the optimal route connecting them. The situation is complicated by the fact that it is possible, in heavily inhomogeneous media, to have multiple rays connecting two points. The principle of Fermat continues to hold, and all rays will be \emph{locally} optimal, but not necessarily \emph{globally} optimal. Mathematical speaking, this implies that the eikonal function has multiple branches. In the context of networking, this means that it is possible to have multiple routes connecting a source-destination pair, each of which achieved a local minimum of cost, but not necessarily the global minimum.
These practical issues are addressed in \cite{catanuto3}.

\end{document}